# NovoMol: Recurrent Neural Network for Orally Bioavailable Drug Design and Validation on PDGFRα Receptor

Ishir Rao [1*]

**Affiliations**

1   Chatham High School – 255 Lafayette Avenue, Chatham, New Jersey, USA

\*   Correspondence: ishirraov@gmail.com

# Abstract

Longer timelines and lower success rates of drug candidates limit the productivity of clinical trials in the pharmaceutical industry. Promising de novo drug design techniques help solve this by exploring a broader chemical space, efficiently generating new molecules, and providing improved therapies. However, optimizing for molecular characteristics found in approved oral drugs remains a challenge, limiting de novo usage. In this work, we propose NovoMol, a novel de novo method using recurrent neural networks to mass-generate drug molecules with high oral bioavailability, increasing clinical trial time efficiency. Molecules were optimized for desirable traits and ranked using the quantitative estimate of drug-likeness (QED). Generated molecules meeting QED's oral bioavailability threshold were used to retrain the neural network, and, after five training cycles, 76% of generated molecules passed this strict threshold and 96% passed the traditionally used Lipinski's Rule of Five. The trained model was then used to generate specific drug candidates for the cancer-related PDGFRα receptor and 44% of generated candidates had better binding affinity than the current state-of-the-art drug, Imatinib (with a receptor binding affinity of -9.4 kcal/mol), and the best-generated candidate at -12.9 kcal/mol. NovoMol provides a time/cost-efficient AI-based de novo method offering promising drug candidates for clinical trials.

# 1. Acknowledgements

We would like to thank Dr. Yelena Naumova from Chatham High School for her support and encouragement throughout the project.

*1.1 Funding*

This research did not receive any specific grant from funding agencies in the public, commercial or not-for-profit sectors.

*1.2 Declaration of Interest*

The authors declare that there are no conflicts of interest.

## 2. Introduction

Drug discovery and development is the first stage of the drug development process, in which drug candidates are identified and sent to clinical trials for further study [1]. This research and development (R&D) process is vital, as it narrows down the molecular space for preclinical researchers from approximately $10^{60}$ molecules to only a small number of promising compounds to be tested further. However, even with countless technological breakthroughs, pharmacological research productivity has been reduced. Current research methods are costly, with an 80-fold increase since 1950 in cost per FDA-approved drug [2]. Expensive research methods result in increased pharmaceutical prices, with an annual increase of 6% in prescription drug prices and over 10% in cancer drug prices in the past five years [3]. The medical administration route (oral, intravenous, etc.) also determines the pricing of drugs, with oral drugs tending to be more cost-effective, along with being non-invasive and convenient [4]. Additionally, these research methods are not easily scalable, with different drug screening techniques such as high throughput screening (HTS) and virtual screening (VS) taking up to 3 months and often generating many false positive hits, leading to high failure rates in the number of viable drug candidates [5] [6]. As a result, only 13.8% of drug candidates pass the first stage of clinical trials, and only 3.4% of oncology candidates pass these trials [7]. Examining the common characteristics of failed candidates, ~50% lack clinical efficacy, ~30% have high toxicity, and ~15% lack drug-like properties. These issues present a clear need for a time and cost-efficient approach to drug discovery that optimizes molecular characteristics and reduces the failure rate of drug candidates by quickly narrowing down the chemical search space.

One promising method is in-silico de novo drug design, which computationally generates potential drug molecules from scratch. Current drug discovery techniques use molecular libraries to find a pre-existing compound with desired pharmacological and physiochemical properties. De novo drug design using machine learning, on the other hand, is not limited by existing molecules and can fully explore the chemical space of viable synthetic molecules. However, efficiently generating drugs with specific molecular characteristics and for specific drug targets has proven to be difficult. Autoencoders, evolutionary algorithms, and other computational methods often fail at either designing viable synthetic molecules or molecules with desired characteristics [8]. Recurrent neural networks (RNNs), on the other hand, have seen some success. By encoding molecules into strings of text through the Simplified Molecular Input Line-Entry System (SMILES), recurrent neural networks can find patterns in SMILES molecules and can be successfully trained to predict SMILES characters [9].

In this work, we propose NovoMol: an evolutionary approach to de novo drug design to optimize for specific drug characteristics using RNN in a cost and time-efficient manner (Fig. 1). We then apply NovoMol to synthesize drug candidates for a specific drug target and analyze the efficacy of this approach.

An RNN generates novel molecules in the format of SMILES strings, and, taking inspiration from evolutionary approaches, the top generated molecules are used to retrain the network through transfer learning. We use a novel measure of drug-likeness known as Quantitative Estimate of Druglikeness (QED) to rank generated drugs based on their molecular properties [10]. The mean QED for FDA-approved orally bioavailable drugs is 0.586. NovoMol optimizes molecules to stay above this threshold, providing an objective-based approach to de novo drug design specifically for orally bioavailable drugs.

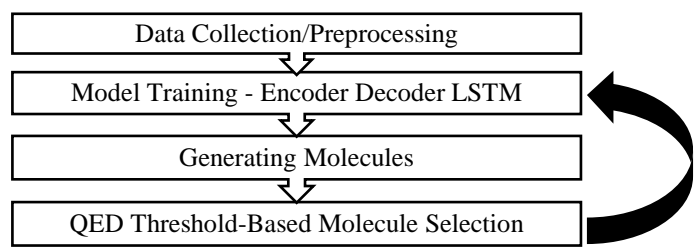

**Figure 1. NovoMol Training Outline:** The general outline of the NovoMol de novo drug design training cycle, which takes inspiration from evolutionary algorithms. NovoMol was trained across multiple iterations, increasing accuracy, and optimizing molecular properties

## 3. Methodology

*3.1 Data Collection/Preprocessing*

We collected 482,000 molecules from ChEMBL, an open-source chemical database manually curated by the European Bioinformatics Institute [11]. Using the ChEMBL Python API, we filtered for drug molecules that have a calculated QED value greater than or equal to 0.586 that are orally administered. Optimizing for oral bioavailability is vital in drug development, since the majority of FDA-approved drugs are administered orally and oral drugs tend to be safer and cheaper. Additionally, drugs lacking oral bioavailability also tend to fail in the clinical stages, making it a key factor to optimize in order to reduce the failure rate.

Molecules were retrieved in SMILES format to make it easy for a neural network to interpret and learn from. SMILES uses ordered characters to represent the overall chemical structure of molecules. Unique characters represent atoms, bonds, and rings, allowing SMILES strings to concisely store the molecular information of our 482,000 molecules which, in total, had 15 million SMILES characters and 43 unique characters.

Finally, we added start (!) and stop (E) characters to each of our SMILES molecules in their respective locations, to allow the model to differentiate between molecules, increasing our unique character count to 45 characters. Since each molecule had a different character length, we padded them with stop characters (E) to a length of 100 characters (Fig. 2). SMILES molecules were one-hot encoded, allowing each character to be represented by a 45-dimensional vector of zeroes with ones in the character index and for the model to easily learn patterns in these retrieved molecules.

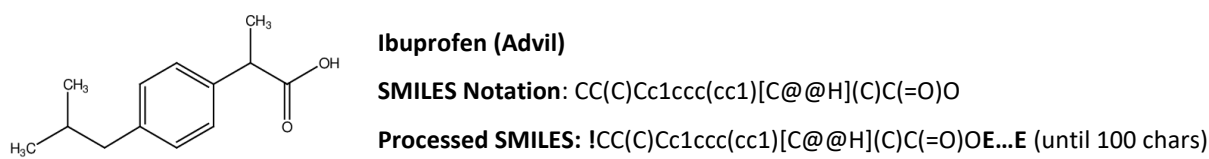

**Ibuprofen (Advil)**

**SMILES Notation**: CC(C)Cc1ccc(cc1)[C@@H](C)C(=O)O

**Processed SMILES: !**CC(C)Cc1ccc(cc1)[C@@H](C)C(=O)O**E...E** (until 100 chars)

**Figure 2. Example SMILES Molecule:** Example SMILES notation for Ibuprofen, along with the processed SMILES molecule which was inputted into the model.

*3.2 Recurrent Neural Networks*

Recurrent Neural Networks (RNNs) have proven successful in capturing sequential and structured data and interpreting meaning, especially in the field of natural language processing. This is achieved by the unique structure of RNNs, which consists of multiple duplicates of the same neural network with each passing selective data to the next through hidden states. Each hidden state is used as the input to the next neural network, allowing each output to be affected based on all hidden states.

The data passed on by these hidden states, however, is diluted over time. This results in the model "forgetting" important information over time which, in problems such as writing sentences or using proper grammar, can result in major inaccuracies with the model. In SMILES molecules, for example, a set of parentheses are used to show branching. However, if the parentheses are opened but not closed, it results in an invalid molecule. Due to the large gap between open and closed parentheses, vanilla RNNs can forget to close branches, leading to significant issues. As a result, we used a long short-term memory (LSTM) recurrent neural network [12].

LSTM networks are a type of RNN specifically designed to solve the "forgetting" problem. LSTMs, unlike a regular RNN, use groupings of three neural networks, known as a cell, instead of just one neural network duplicated. The three neural networks are the forget gate, input gate, and output gate and they each have a specific role to determine what information is passed forward to the next cell along with what information needs to be stored in an additional "long-term storage" cell state. By using both a hidden state and a cell state, the LSTM can remember key details in long-term memory. This allowed our model to be much more accurate since, due to the nature of SMILES molecules, even a single wrong character while generating molecules can create the wrong molecule.

Our overall neural network used an encoder-decoder structure consisting of an LSTM, input, and dense layers using Python and the TensorFlow and Keras libraries [13] (Fig. 3). An encoder turns the input (which in our case is each character of our 100-character long SMILES molecules) into a latent representation, which reduces the dimensionality of the data while still trying to convey the relevant information of the molecule. The decoder then reads this compressed format of our original SMILES molecule and tries to recreate the original molecule. Initially, the difference between the original molecule and the molecule outputted by the decoder, which is known as reconstruction error, will be significant. During training, both the encoder and decoder work toward finding the most effective way the encoder can compress the data and the decoder can recreate the molecule.

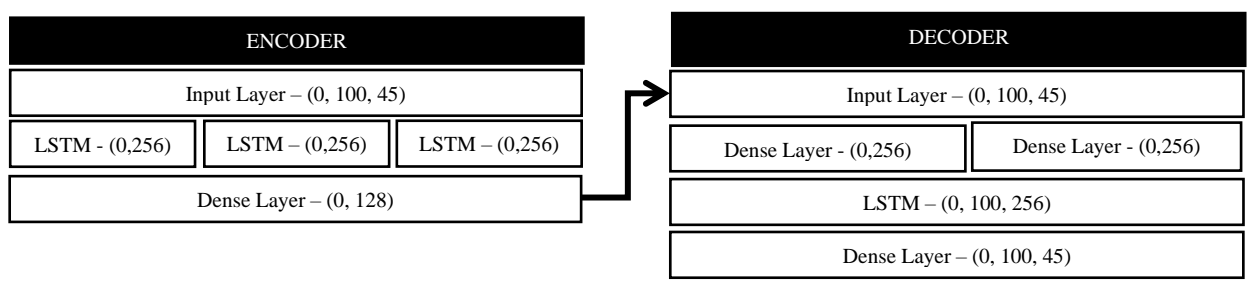

**Figure 3. Model Architecture:** The model architecture of our encoder-decoder model, along with the outputs of each layer. Side-by-side layers are stacked layers. The output of the encoder is the input to the decoder.

*3.3 Training NovoMol*

By training our complete model on the 482,000 molecules that we have retrieved, it learned to effectively compress and then recreate a molecule with minimal reconstruction error. Our encoding layer was composed of an input layer of size 100 (the length of our SMILES molecules), three LSTM layers, each of size 256, and a dense layer of size 128, which condensed the hidden and cell states from our LSTM. The decoder consisted of an input layer, two dense layers for our two states, one LSTM layer of size 256, and finally a dense layer that yielded output logits and was converted to probabilities using a SoftMax layer. This resulted in 761,773 trainable parameters. Backpropagation trained the network using the Adam optimizer and categorical cross-entropy loss. SMILES molecules 100-time steps (characters) were input into the network in batches of size 256, and the model was trained for 225 epochs on Google Colab Tesla P100 GPU [14], yielding a 99.96% training accuracy and 99.86% test accuracy.

To generate SMILES molecules, an intermediary model was added between the encoder and decoder. This model decoded the latent space generated by our encoder into LSTM hidden and cell states to be inputted into the decoder model. By separating the base model into two standalone models and adding an intermediary model between them, the model was able to predict molecules one character at a time, allowing us to generate novel molecules.

By first providing a start character (!) to our model, our model was able to predict the rest of the characters in a SMILES molecule. However, since our accuracy was high, it was likely that the molecules created would be the molecules present in our training set. As a result, we altered the temperature (T) of the SoftMax activation function, adding diversity to the generated molecules. When $T > 1$, there was more diversity/mutation in our molecules, but also worse-quality molecules. When $T < 1$, the model was more conservative with the generated molecules. For each generated molecule, we randomly selected a temperature between 0.75 and 1.25 to allow for a diverse set of generated molecules.

*3.4 Quantitative Estimate of Drug-Likeness (QED)*

During drug discovery, researchers need methods of quickly evaluating and narrowing down possible drug candidates to reduce costs and increase efficiency. Many guidelines, such as Lipinski's Rule of Five (Ro5), have been used to identify ideal candidates for further research. These guidelines are pass or fail and look at multiple molecular characteristics to provide an approximation of whether candidates are likely suitable drugs. A molecule passing Rule of Five is defined as having a molecular weight < 500 Daltons, cLogP < 5, H-bond donor < 5, and H-bond acceptor < 10.

However, this pass/fail system is not extremely accurate, with more than 16% of approved oral drugs failing Ro5. As a result, a quantitative metric for drug-likeness would provide a much more accurate and versatile method for researchers to compare drug candidates. The Quantitative Estimate of Drug-Likeness, as defined by Equation 1, considers eight molecular characteristics, which are weighted (Table 1) based on trends seen in FDA-approved drugs. Additionally, a QED value of 0.37 or above indicates that the molecule would also pass Ro5 and other prominent guidelines. The QED of a valid SMILES molecule can be calculated using RDKit, a Python module for cheminformatics [15], and ranges from 0 to 1, with 1 being the molecule with the most favorable properties. Higher QED indicates that a drug candidate would be more likely to receive FDA approval. A summary of QED thresholds used in this work can be found in Table 2.

**Equation 1. QED Calculation:** The weighted QED calculation, where *d* is each molecular characteristic value, *w* is the weight applied to each, and *n* is the number of molecular characteristics.

$$QED_w = \exp\left(\frac{\sum_{i=1}^{n} w_i \ln d_i}{\sum_{i=1}^{n} w_i}\right)$$

| Molecular Weight | ALOGP | HBD | HBA | PSA | ROTB | AROM | ALERTS |
|---|---|---|---|---|---|---|---|
| 0.55 | 0.46 | 0.61 | 0.05 | 0.06 | 0.65 | 0.48 | 0.95 |

**Table 1. Weighted Molecular Descriptors:** Optimized weightings for each molecular characteristic [10]

| QED Value | Indication |
|---|---|
| > = 0.37 | The molecule would pass Ro5, Ghose, Veber, and Gleeson guidelines |
| > = 0.586 | Calculated mean QED of FDA-approved oral drugs found in ChEMBL – higher QED results in more favorable oral drug properties. The oral bioavailability threshold of 0.586 was calculated by retrieving all FDA-approved oral drugs from ChEMBL and calculating the mean QED from that set using RDKit. |

**Table 2. QED Thresholds:** Calculated QED thresholds used [10]

*3.5 Transfer Learning*

Taking inspiration from the success of evolutionary algorithms in drug design, we also implement transfer learning to create multiple generations of our model. Transfer learning (TL) is extremely useful in this problem since the goal of our model is very specific and lacks significant data. In TL, an original model is trained on one task and then re-purposed to be used on a related task. Since the original model already has pre-existing knowledge about a similar task, it is often more accurate and requires less data to train. To create a model which can consistently generate valid SMILES molecules with high QED, we can apply transfer learning multiple times. We duplicate the original model and retrain it to focus on the chemical space with higher QED molecules. The next-generation model is trained on the molecules generated by the previous generation with a QED value above our oral bioavailability threshold of 0.586. In total, we ran five generations of the model to increase model performance.

# 4. Results

*4.1 De Novo Generation*

After only the first generation of the model, 84.23% of the molecules passed the traditionally used Lipinski's Rule of Five and 51.82% passed the oral bioavailability QED threshold of 0.586. To evaluate whether the novel generated molecules were in the same general chemical space as the training dataset, we applied principal component analysis (PCA) to visualize the eight molecular descriptors used to define QED. We compared the molecules generated by the first-generation model with the original training dataset using PCA in Fig. 4 and visualized each property individually in Fig. 5. PCA reduces the dimensionality of this data to two dimensions for plotting, and the significant overlap of generated molecules with training molecules indicates that the generated molecules occupied a similar chemical space as the training set, but the molecules were not copied.

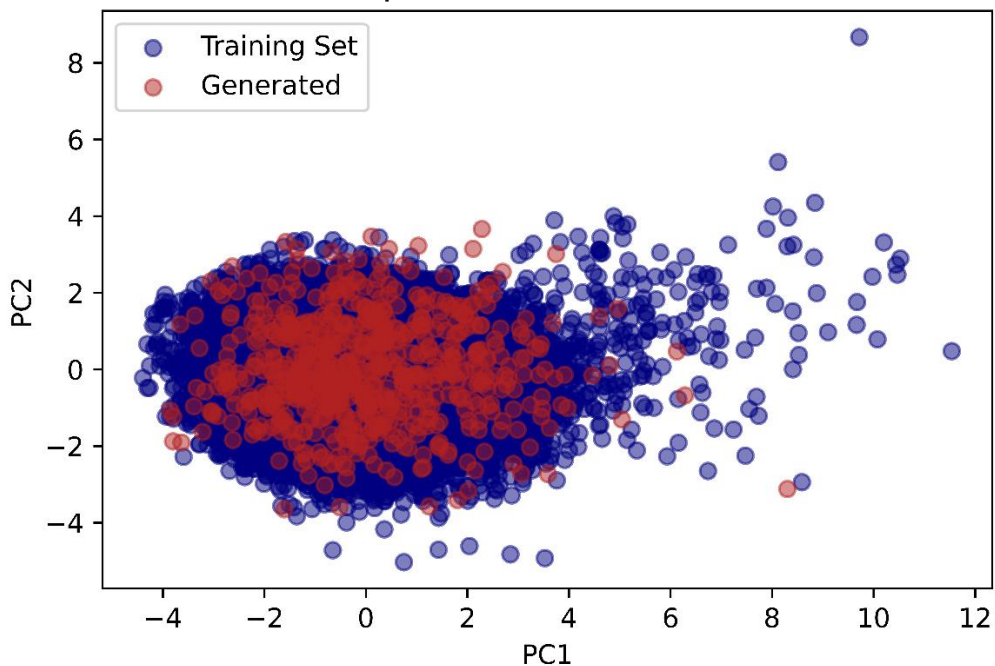

**Figure 4. Principal Component Analysis of first-generation molecules:** Principal Component Analysis (PCA) of the molecular descriptors used to calculate QED in the first-generation molecules and the original training set. Eight molecular descriptors were calculated for each generated molecule and 20,000 randomly selected molecules from the training set.

In Fig. 5, the properties overlap significantly, which signified our model can created novel molecules that are similar to the training data. However, we also saw that some of our generated molecules have structural alerts (ALERTS), indicating potential toxicity and adverse drug reactions to our molecules. This poses a potential error in NovoMol drug generation capabilities since it did not learn to avoid possible toxicity while designing drugs. However, future generations and further training may solve this problem, and this can easily be caught by running structural alert algorithms such as the one we used in RDKit.

We also calculated the structural diversity of the generated molecules. Having a variety of structurally diverse drug candidates for redundancy during drug testing and approval is important, since many candidates may fail throughout the process. Using RDKit, we converted our SMILES molecules into Morgan fingerprints, which gives us a way to represent the structure of the molecule. From these fingerprints, we calculated the Tanimoto similarity (*S*) for each molecule pair and found the average *S*. Tanimoto similarity ranges between 0 and 1, with 0 being the most structurally diverse pair. *S* can be calculated for all pairs of molecules *x* and *y*, where $|m_x \cap m_y|$ is the common fingerprints (similar/same structure) and $|m_x \cup m_y|$ is the total number of Morgan fingerprints by [16]:

$$S(x,y) = \frac{|m_x \cap m_y|}{|m_x \cup m_y|}$$

The average Tanimoto similarity of 2,000 randomly selected molecules from the training set was 0.151. The novel-generated molecules in the first-generation model had an average of 0.114, which shows that there is significant structural diversity. The generated molecules were more structurally diverse than the training dataset, which is beneficial for drug development.

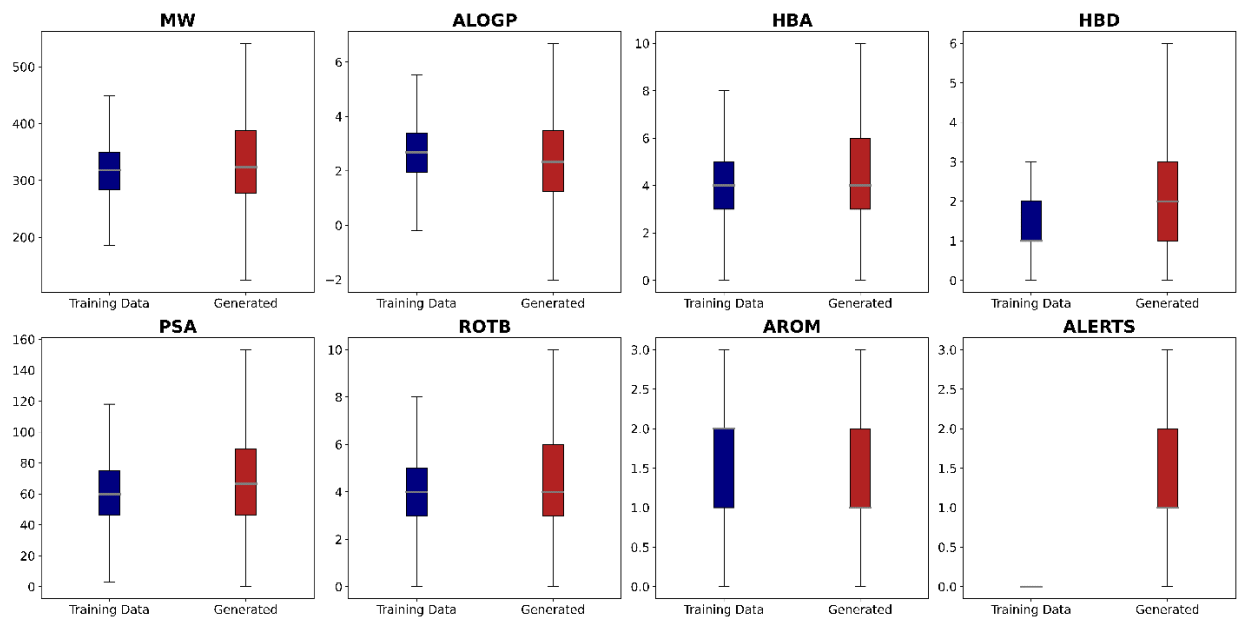

**Figure 5. Molecular descriptor comparison with initial training data:** Box plots of each of the molecular descriptors (molecular weight, AlogP, H-Bond acceptors, H-Bond donors, polar surface area, rotatable bonds, aromatic rings, and structural alerts) in the training data and the generated data. The median and lower quartile values are equal in the distribution of aromatic rings, so we have not added a median mark. The training data has 0 structural alerts, and our generated data has a median of 1 structural alert.

*4.2 Evolutionary Optimization*

We ran four iterations of transfer learning on the original NovoMol model, resulting in five total generations of the model to further improve the results. 250k SMILES characters were generated in total, resulting in 7518 valid novel molecules, all of which are unique from each other and the initial training dataset. Of those molecules, 92.99% pass Lipinski's Rule of Five, and 69.76% pass the QED oral bioavailability threshold of 0.586. Only 0.4% of the molecules generated were invalid molecules. Each molecule was, on average, 33 SMILES characters long.

The generated molecules were plotted using PCA across all generations and compared them to the original training data as well (Fig. 6). We can observe that, across iterations, the model closes in on a more optimal part of the chemical space with the best molecular characteristics for high QED. We also visualized each molecular property again, comparing the distributions before and after transfer learning (Fig.7). We can see that, for most properties, there is still significant overlap. However, for some properties (AROM, HBD, ROTB), the model after five iterations has found a more optimal chemical space by reducing the values. Additionally, by the distribution of ALERTS before and after, it is shown that NovoMol learned to design potentially non-toxic molecules more consistently.

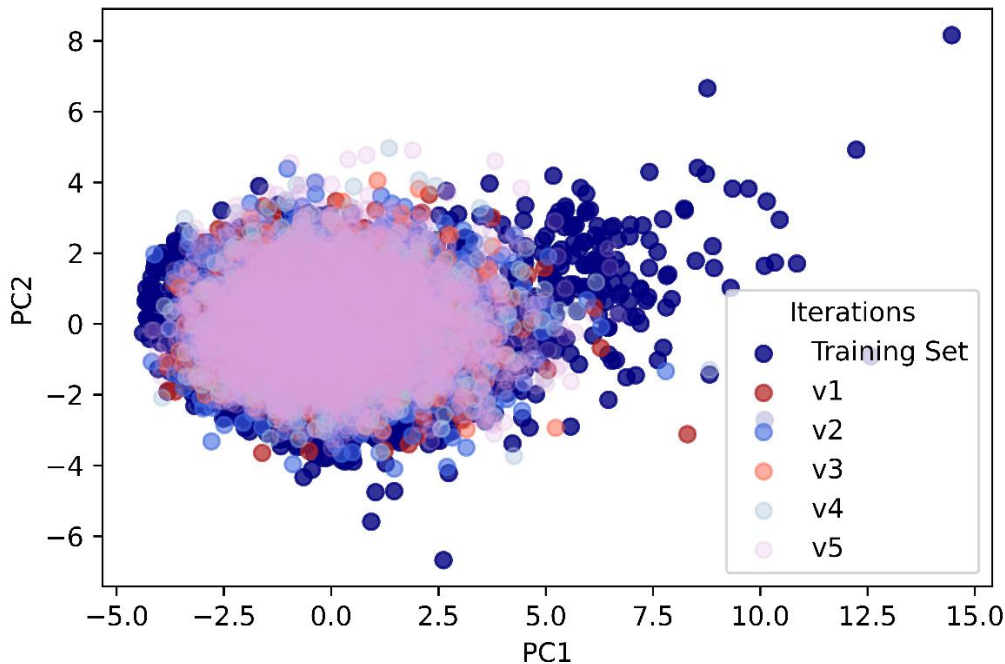

**Figure 6. Principal Component Analysis of Molecules:** Principal Component Analysis (PCA) of the molecular descriptors used to calculate QED across the five generations and the original training set. Eight molecular descriptors were calculated for each of the 6833 generated molecules and 20,000 randomly selected molecules from the training set

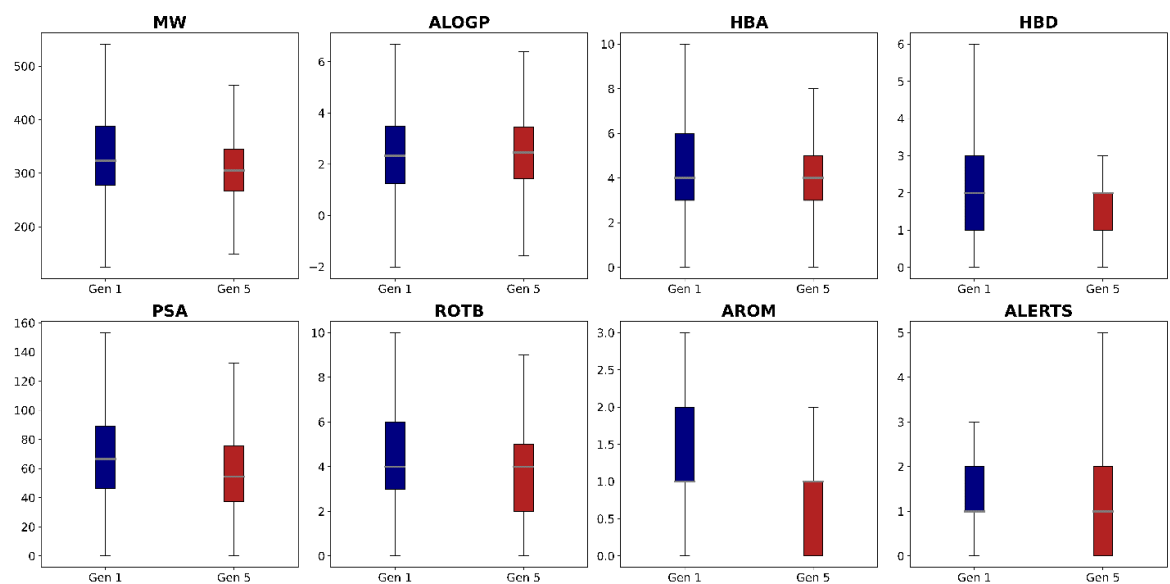

**Figure 7. Comparison of generated molecular descriptors:** Box plots of each of the molecular descriptors used to calculate QED before transfer learning and after. We see significant overlap in the distributions of both, except in the number of aromatic rings

We can analyze the percentage of generated molecules passing each QED threshold to determine whether NovoMol learned to optimize molecules across generations. We can see a nearly 150% increase in the percentage of molecules passing the QED oral bioavailability threshold of 0.586, as shown in Fig. 8 and Table 3. As expected, the percentage that passes the Rule of Five is much higher than those that pass the oral bioavailability threshold, as QED is significantly stricter than the Rule of Five. After five generations, we see the percentage level off, with the highest percentage in the fourth generation at 76%. Additionally, in the fourth generation, 96% of molecules pass the traditional Lipinski's Rule of Five. This is a significant improvement from previous works, such as by Yasonik [17], which, after five iterations of training an LSTM network, reached a 32.62% pass rate of the Rule of Five for generated molecules. As a result, our work is nearly a 3-fold improvement in drug design accuracy. We can also see QED optimization by the spread of QED values in Fig. 9, where, after 5 iterations, most molecules are generated with a QED around 0.79, indicating the model optimizes for higher QED.

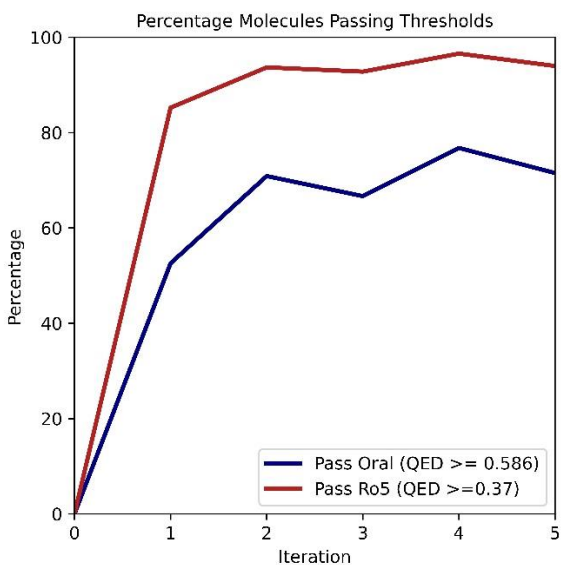

**Figure 8. Molecule benchmark percentages across generations:** Lines showing the percentage of molecules passing our two QED thresholds. Iteration 0 represents molecules generated randomly, and not molecules generated by an iteration of the model. We generated 10,000 molecules by randomly selecting SMILES characters for iteration 0, and none passed either threshold.

| Iterations | Pass Rule of Five and other traditional guidelines, with a QED >= 0.37 (%) | Pass the QED oral bioavailability threshold of QED >= 0.586 (%) |
|---|---|---|
| 1 | 84.23 | 51.82 |
| 2 | 93.36 | 70.56 |
| 3 | 91.96 | 66.07 |
| 4 | 96.13 | 76.43 |
| 5 | 93.72 | 71.33 |

**Table 3. Molecular quality percentage progression:** Percentages of generated molecules passing thresholds at each iteration

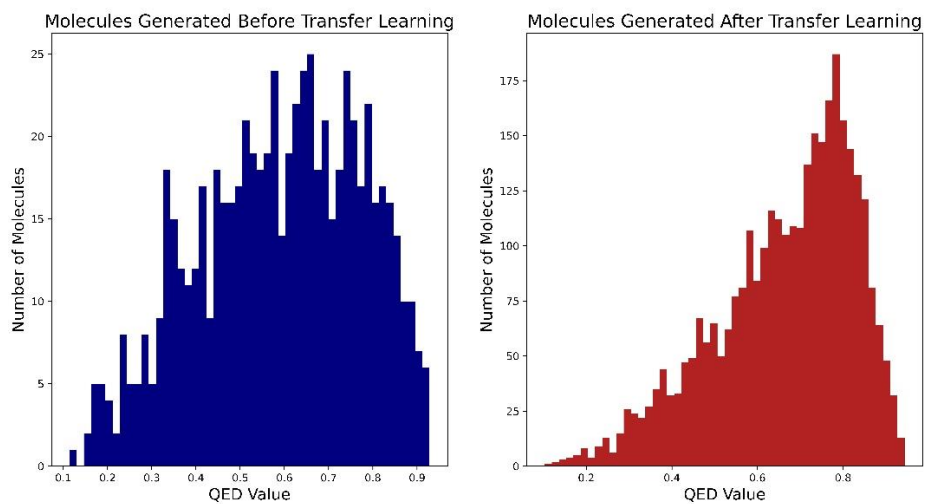

**Figure 9. Molecular Quality Comparison:** Histograms showing the spread of QED values of generated molecules before and after transfer learning. After transfer learning, the QED of molecules were skewed higher.

Along with QED, general model accuracy increased with transfer learning. Before transfer learning, 1.17% of generated molecules were invalid. After, only 0.27% of molecules were invalid. Additionally, the structural diversity of the molecules increased, with the Tanimoto similarity decreasing from 0.114 to 0.109. This indicates that, while the molecules are closing in on an optimal chemical space, the model is still able to generate diverse molecules which can prove useful in drug development where structure-related failures of drug candidates are common.

## 5. Discussion

To test the efficacy of NovoMol, we can apply the model to design drugs for a specific drug target. In our work, we chose to design drugs to inhibit mutated platelet-derived growth factor receptor alpha, also known as PDGFRα [18]. This is a membrane-bound tyrosine-kinase receptor, and, when mutated, plays a significant role in the increased growth of cancer cells. When the ligand, PDGF, activates the non-mutated receptor, the receptor is phosphorylated and signaling pathways encourage mitotic activity in cells, leading to an increase in cell division. However, when the receptor is mutated, the extracellular domain is replaced with oncoprotein TEL ETV-6. Rather than waiting for the PDGF ligand, the mutated receptor phosphorylated without any extracellular ligand, in a process known as self-NovoMol: Recurrent Neural Network for Orally Bioavailable Drug Designdimerization. As a result, the receptor is constantly signaling the cell to divide, with or without the presence of PDGF. In a cell with mutated DNA, this can lead to cancerous cells and the growth of tumors.

PDGFRα has been found in gastrointestinal stromal tumors (GISTs) and chronic eosinophilic leukemia, along with other benign and malignant neoplastic diseases [19]. These cancers can be treated through chemotherapy, radiation, and through target-specific drugs [20]. Oral cancer drugs, such as Gleevec (imatinib), competitively inhibit the PDGFRα ATP-binding domain and can prevent or stop the growth of cancer cells and reduce the size of tumors non-invasively [21]. Approximately 39-45% of patients treated with imatinib, which is the current state-of-the-art drug, can remain in remission for multiple years. However, imatinib also has several side effects which can impact the quality of life and can turn serious quickly. As a result, we must design more effective drugs with less risk of side effects to treat cancers safely and more efficiently.

*5.1 Transfer Learning for Tyrosine-Kinase Inhibitors*

To use NovoMol to computationally generate inhibitors for PDGFRα, we need to retrain our model using transfer learning. We collected a dataset of 450 known inhibitors from ChEMBL and fed the fifth-generation model with these SMILES molecules. By doing so, NovoMol learned the molecular patterns found in these inhibitors, on top of the existing knowledge it had about oral drugs and optimizing for high QED. This is a technique known as Ligand-Based Drug Design which allows specific novel molecules to be designed with minimal training based on existing compounds. After training, we generated 311 potential drug candidates.

*5.2 Virtual Screening of Generated Molecules*

To virtually screen these drug candidates and determine whether NovoMol was able to design inhibitors to PDGFRα, we can determine the binding affinities between the generated candidates and the receptor at the protein-ligand complex. Binding affinity is the measure of the strength of interaction between two molecules and can help us determine whether a molecule could bind to the receptor and inhibit its function. Binding affinity is measured in kcal/mol, where a lower value results in a higher binding affinity [22]. It indicates how much of a drug is required for an intended effect on the drug target (PDGFRα), so when our binding affinity is high, that also means that less drug is required. Finally, with a lower dosage, we can also expect lowered side effects as a result of the drug. Imatinib has a binding affinity of -9.4 kcal/mol, so if our drug candidates have a lower kcal/mol, we can infer that they would most likely require less dosage and have fewer side effects than imatinib.

To calculate binding affinity, we first used Open Babel and PyMOL [23], which are open-source molecular visualization systems, to translate our generated SMILES molecules into the appropriate file format for binding and add polar hydrogens to find hydrogen bond interactions and Van der Waals forces. Then, we download the structure of PDGFRα from the RCSB Protein Data Bank, convert it into the appropriate file format, and determine the docking analysis search space of the protein in PyMOL.

We use AutoDock Vina, a modeling simulation software for protein-ligand docking from the Scripps Research Institute [24] [25], to screen our molecules and determine their binding affinities and best binding locations. We used this to conduct a batch virtual screening, with a search exhaustiveness value of 32 to determine the best positioning of each molecule with the highest binding affinity to the receptor. Then, for all molecules with a binding affinity lower than -9.4 kcal/mol, we used Vina again with an exhaustiveness of 256 for high levels of accuracy.

*5.3 PDGFRα Inhibitors*

We tested these generated drug candidates for their QED values and found that 49.5% passed the Ro5 and 16.4% (51 molecules) passed the oral bioavailability threshold. This demonstrated an improvement from the initial dataset provided to the model, which had only 14.2% of molecules passing the oral bioavailability threshold. This indicates that NovoMol used the pre-existing knowledge of optimizing for QED while designing these drug candidates to make them more favorable candidates.

Out of our 311 generated candidates, 44% (137 molecules) had a better binding affinity than imatinib on the PDGFRα receptor. Additionally, the majority of the generated candidates with a kcal/mol lower than -9.4 were also passed the QED oral bioavailability threshold. In Figure 10, we can see the PDGFRα receptor, and all 311 generated molecules at their best binding locations on the receptor. Imatinib and the best molecule bind to an active site. This can be assumed since imatinib is a competitive inhibitor for the ATP-binding domain on the receptor, blocking the active site on the right side of Figure 10b.

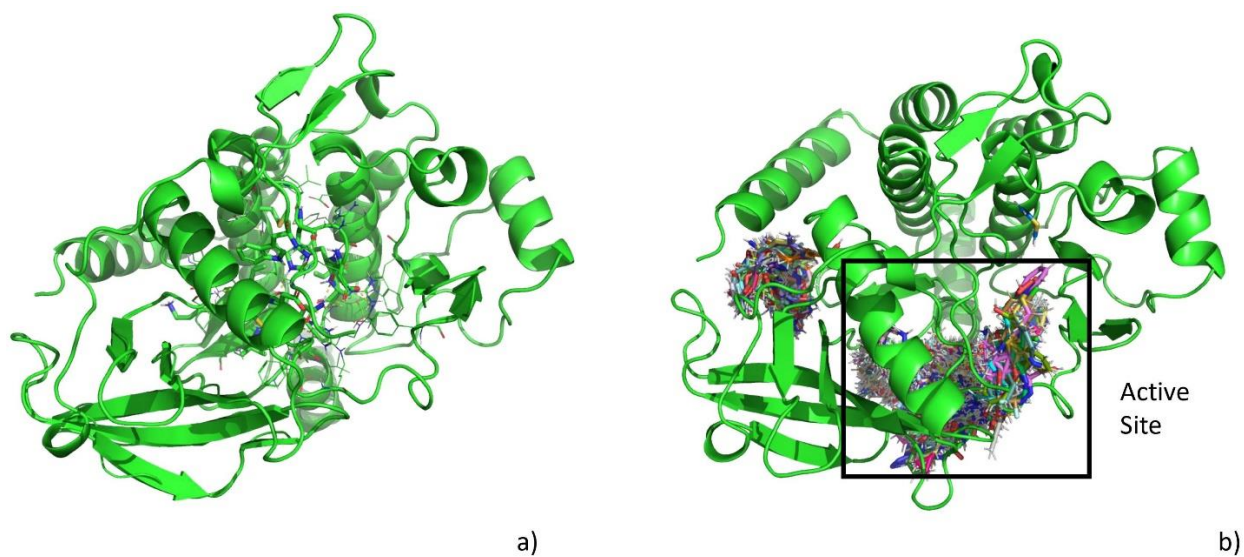

**Figure 10. Protein structure of PDGFRa:** Figures downloaded from PyMOL. (**A**) The PDGFRα receptor, downloaded from the RCSB Protein Data Bank. (**B**) The calculated AutoDock Vina binding poses for all 311 molecules on the receptor. We can locate two distinct sites on the receptor, one which is a known active site due to the binding location of Imatinib

In Table 4, we found the five generated molecules with the best binding affinity that were also orally bioavailable, and we compared them to imatinib. Out of the 311 generated candidates, 5% (16 molecules) are both orally bioavailable and have a better binding affinity than Imatinib. Based on these results, we find that not only was NovoMol able to generate inhibitors for the PDGFRα receptor, but it was also able to optimize the QED of these molecules for oral bioavailability.

**Comparison of PDGFRa-Targeted Drugs**

| QED Value | Binding Affinity (kcal/mol) | Protein Binding |
|---|---|---|
| **0.389** | -9.4 | 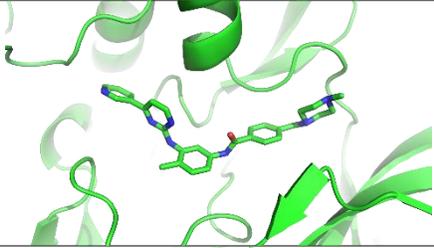 |
| **0.686** | -12.9 | 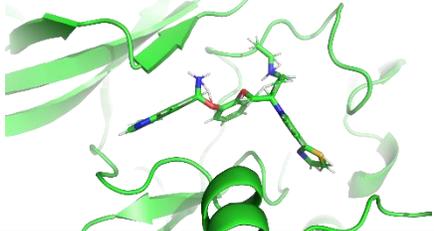 |
| **0.609** | -12.2 | 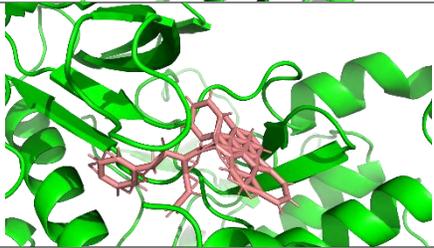 |
| **0.596** | -12.1 | 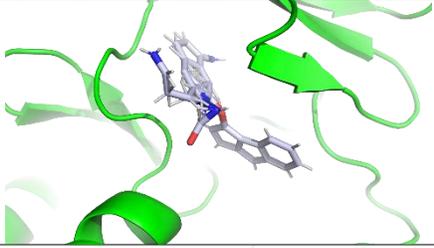 |
| **0.622** | -12.0 | 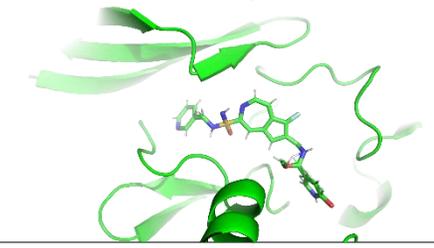 |
| **0.606** | -10.9 | 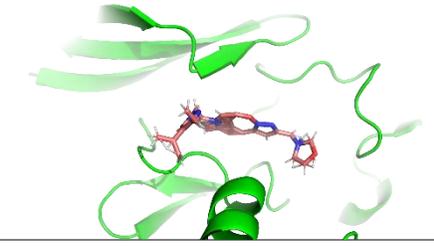 |

**Table 4. Comparison of PDGFRa-Targeted Drugs:** The current state-of-the-art chemotherapeutic drug, imatinib, along with the top five generated molecules by NovoMol. We first filtered the 311 molecules for only those which passed the oral bioavailability threshold of 0.586. Then, we ranked them based on binding affinity, starting with the lowest kcal/mol (highest binding affinity). Finally, we show their calculated binding poses to PDGFRα.

## 6. Conclusion

In this work, we applied an encoder-decoder model along with a novel metric for drug-likeness to create *NovoMol: an effective method for orally bioavailable de novo drug design*. First, we created a long short-term memory (LSTM) recurrent neural network, which we trained on existing molecules with a high quantitative estimate of drug-likeness (QED). Molecules with high QED tend to have more favorable properties for oral bioavailability. The model was able to successfully generate novel and diverse molecules with similar properties. Then, we fine-tuned the model through an iterative approach, where we retrained the model on the best-generated molecules to optimize molecular properties for high QED. After five iterations, there were significant improvements in the molecules generated, as seen by the percentages of molecules passing two thresholds across iterations. The final iteration had molecules with much higher QED on average, indicating that the model effectively learned to generate orally bioavailable drugs.

To validate the efficacy of NovoMol, we retrained our model to generate drugs for the oncoprotein PDGFRα, to inhibit the protein and the growth of cancerous cells. Existing drugs have significant side effects due to dosage amounts. We were able to design drugs with more favorable drug properties (higher QED) and require less dosage (higher binding affinity), resulting in suitable drug candidates that would likely reduce side effects. While NovoMol does offer promising candidates, significant laboratory testing is required to confirm these properties and test for other effects. This demonstrates the application of NovoMol to design effective drugs with limited access to data for specific targets.

There are three primary benefits to the usage of NovoMol. First, this approach to de novo drug design quickly mass generates potential drug candidates, optimizing for vital molecular characteristics that increase the probability of the drug candidate being successful. Second, the model does not require significant amounts of data to generate new candidates for a specific target. The latest iteration of the model can easily be retrained and repurposed for a specific application, allowing it to be extremely versatile and commercially useful. As a result, it has universal applications and can design for any target in a matter of minutes. Finally, NovoMol taps into the complete chemical space, rather than being restricted by existing molecular libraries. As a result, it could potentially offer novel and improved therapies.

Future improvements to NovoMol include using other types of AI such as reinforcement learning, along with using novel scoring systems to encourage the model to generate increasingly diverse and consistently favorable molecules, molecule generation through reaction-based de novo design and extending the model for virtual screening through property prediction using machine learning methods and structural alerts [26] [27]. Additionally, testing for several other drug targets will help validate NovoMol's efficacy. Some potential targets include AChe (Alzheimer's disease), HIV-1 RT (HIV/AIDS), HER2 (breast cancer), and EGFR (lung cancer).

Researchers in drug development pipelines are increasingly opting to use in-silico de novo drug design approaches. A versatile and accurate approach such as NovoMol could help reduce the costs and timelines limiting drug development currently. By generating better-quality drug candidates, we can accelerate the development of novel therapies and introduce these vital drugs to the market sooner.

## 7. Declarations

*7.1 Ethical Approval*

No humans and/or animals were studied in this work, so this declaration is not applicable.

*7.2 Funding*

No funding was received for this research, so this declaration is not applicable.

*7.3 Availability of data and materials*

The dataset supporting the conclusions of this article are available in the NovoMol repository in Python and run on Google Colaboratory at this source: https://github.com/ishirraov/NovoMol. Molecular data, in SMILES format, was extracted from the open-source ChEMBL Dataset [11].

## 8. Bibliography

[1] Commissioner, O. of the. *The drug development process*. U.S. Food and Drug Administration. (2020).

[2] Scannell, J., Blanckley, A., Boldon, H. *et al.* Diagnosing the decline in pharmaceutical R&D efficiency. *Nat Rev Drug Discov* **11**, 191–200 (2012). https://doi.org/10.1038/nrd3681

[3] Vincent Rajkumar, S. The high cost of prescription drugs: causes and solutions. *Blood Cancer J.* **10**, 71 (2020).

[4] Kim J, De Jesus O. Medication Routes of Administration. [Updated 2022 Nov 26]. In: StatPearls [Internet]. Treasure Island (FL): StatPearls Publishing; 2022 Jan

[5] Macarron, R., Banks, M., Bojanic, D. *et al.* Impact of high-throughput screening in biomedical research. *Nat Rev Drug Discov* **10**, 188–195 (2011)

[6] Mukhlesur Rahman, Chapter 23 - Antimicrobial Secondary Metabolites—Extraction, Isolation, Identification, and Bioassay, Editor(s): Pulok K. Mukherjee, Evidence-Based Validation of Herbal Medicine, Elsevier, 2015, Pages 495-513

[7] Wong CH, Siah KW, Lo AW (2018) Estimation of clinical trial success rates and related parameters. Biostatistics 20(2):273–286

[8] Mouchlis VD, Afantitis A, Serra A, Fratello M, Papadiamantis AG, Aidinis V, Lynch I, Greco D, Melagraki G. Advances in de Novo Drug Design: From Conventional to Machine Learning Methods. Int J Mol Sci. 2021 Feb 7;22(4):1676

[9] Suresh N, Chinnakonda Ashok Kumar N, Subramanian S, Srinivasa G. Memory augmented recurrent neural networks for de-novo drug design. PLoS One. 2022 Jun 23;17(6)

[10] Bickerton GR, Paolini GV, Besnard J, Muresan S, Hopkins AL. Quantifying the chemical beauty of drugs. Nat Chem. 2012 Jan 24;4(2):90-8

[dataset] [11] Gaulton A, Hersey A, Nowotka M, Bento AP, Chambers J, Mendez D, Mutowo P, Atkinson F, Bellis LJ, Cibrián-Uhalte E, Davies M, Dedman N, Karlsson A, Magariños MP, Overington JP, Papadatos G, Smit I, Leach AR. (2017) 'The ChEMBL database in 2017.' Nucleic Acids Res., 45(D1) D945-D954.

[12] Gupta A, Müller AT, Huisman BJH, Fuchs JA, Schneider P, Schneider G. Generative Recurrent Networks for De Novo Drug Design. Mol Inform. 2018 Jan;37(1-2):1700111

[13] Martín Abadi, et.al TensorFlow: Large-scale machine learning on heterogeneous systems,2015. Software available from tensorflow.org.

[14] Bisong, E. (2019). Google Colaboratory. In: Building Machine Learning and Deep Learning Models on Google Cloud Platform. Apress, Berkeley, CA

[15] RDKit: Open-Source Cheminformatics. https://www.rdkit.org. Accessed 15 Jan 2019

[16] Bajusz D, Racz A, Heberger K (2015) Why is tanimoto index an appropriate choice for fingerprint-based similarity calculations. J Cheminform. 7(20)

[17] Yasonik, J. Multiobjective de novo drug design with recurrent neural networks and nondominated sorting. *J Cheminform* **12**, 14 (2020)

[18] Carvalho, I., Milanezi, F., Martins, A. *et al.* Overexpression of platelet-derived growth factor receptor α in breast cancer is associated with tumour progression. *Breast Cancer Res* **7**, R788 (2005)

[19] Boonstra PA, Gietema JA, Suurmeijer AJH, Groves MR, de Assis Batista F, Schuuring E, Reyners AKL. Tyrosine kinase inhibitor sensitive PDGFRA mutations in GIST: Two cases and review of the literature. Oncotarget. 2017 Nov 26;8(65):109836-109847

[20] Wang F, Lv H, Zhao B, Zhou L, Wang S, Luo J, Liu J, Shang P. Iron and leukemia: new insights for future treatments. J Exp Clin Cancer Res. 2019 Sep 13;38(1):406

[21] Sacha T. Imatinib in chronic myeloid leukemia: an overview. Mediterr J Hematol Infect Dis. 2014 Jan 2;6(1):e2014007

[22] Du X, Li Y, Xia YL, Ai SM, Liang J, Sang P, Ji XL, Liu SQ. Insights into Protein-Ligand Interactions: Mechanisms, Models, and Methods. Int J Mol Sci. 2016 Jan 26;17(2):144

[23] O'Boyle, N.M., Banck, M., James, C.A. *et al.* Open Babel: An open chemical toolbox. *J Cheminform* **3**, 33 (2011).

[24] Eberhardt, J., Santos-Martins, D., Tillack, A.F., Forli, S. (2021). AutoDock Vina 1.2.0: New Docking Methods, Expanded Force Field, and Python Bindings. Journal of Chemical Information and Modeling.

[25] Trott, O., & Olson, A. J. (2010). AutoDock Vina: improving the speed and accuracy of docking with a new scoring function, efficient optimization, and multithreading. Journal of computational chemistry, 31(2), 455-461

[26] Liu, Xuhan, Adriaan P. IJzerman and G. V. van Westen. "Computational Approaches for De Novo Drug Design: Past, Present, and Future." *Methods in molecular biology* 2190 (2021): 139-165

[27] Joshua Meyers, Benedek Fabian, Nathan Brown, De novo molecular design and generative models, Drug Discovery Today, Volume 26, Issue 11, 2021, Pages 2707-2715